# On The Choice Between A Sale-Leaseback And Debt.

**Michael C. Nwogugu**
Address: Enugu 400007, Enugu State, Nigeria.
Email: Mcn2225@aol.com; mcn2225@gmail.com.
Phone: 234 814 906 2100.

**Abstract[1].**
    This article introduces decision models for commercial real estate leasing. The concepts and models developed in the article can also be applied to equipment leasing and other types of leasing.

## 1. Introduction.

Real estate leases are complex long-term contracts requiring simultaneous, continuous and phased performance, and different types of monetary and non-monetary performance by typically unrelated parties. The lessee's propensity to comply with lease terms at specific times is greatly influenced by economic conditions, lessee's resources, and the various costs that may be incurred by the lessee and lessor upon breach of the lease agreement. (Benjamin, Jud & Winkler, 2000). (McNally, Klein & Abrams, 2001); Pretorius, Walker & Chau (2003); (Triantis & LoPucki, 1994); Hylton (1993); Hylton (2002); Michael (2000). Mooradian & Yang (2002); Heyes, Rickman & Tzavara (2004); Katz (1990); Triantis (1993). The typical lease provides the lessor with periodic (quarterly or semi-annual) property inspection rights in order to monitor property conditions.

## 2. Existing Literature On Leases And Sale Lease-Backs.

The existing literature on leasing in the commercial real estate industry is extensive, but the materials don't analyze some of the following issues:

---

1. The optimal conditions for a lease.

2. The optimal lease, and the optimal Rent.

3. The effect of 'incompleteness' of leases on economics of such leases.

4. The choice between leasing and borrowing.

5. The choice between sale-leaseback and no-action, or borrowing.

6. The analysis of commercial property leases as part of the supply chain for retailers and medium/large companies. Location is crucial for retailers. Real estate rents often accounts for more than 30% of the operating expenses of retailers; and more than 15% of operating expenses of other types of companies.

7. The analysis of commercial leasing as a dynamical system.

8. Analysis of commercial property leases as Take-Or-Pay contracts

In the US Sale-leaseback transactions have been the subject of significant litigation about: a) the rights and obligations of lessees and lessors,[2] b) whether sale-leaseback transactions are debt financing

---

*Control*, 5(4), 326-329.

[2] **See:** Murray J (2001). *Off Balance Sheet Financing: Synthetic Leases*. (citing cases). First American Title Insurance Company. http://www.firstam.com/content.cfm?id=4320. Unocal Corp. v. Kaabipour, 177 F.3d 755 (9th Cir. 1999). Some courts have used a fact-based approach and the "economic realities test" – see: *Lazisky*, 72 T.C. at 500-02; *Rich Hill Ins. Agency*, 58 T.C. at 617-19; *Transamerica Corp. v. United States*, 7 Cl. Ct. 441 (1985); *Torez v. Commissioner*, 88 T.C. 702, 721-27 (1987) (sale/leaseback); *Estate of Thomas v. Commissioner*, 84 T.C. 412, 436 (sale leaseback); Rev. Rul. 55-540, 1955-2 C.B. 39 (classification of a lease); Rev. Rul. 68-590, 1968-2 C.B. 66; Rev. Rul. 72-543, 1972-2 C.B. 87. See: Treas. Reg. § 1.1245-1(a)(3) (as amended in 1997); Tech. Adv. Mem. 93-07-002 (Oct. 5, 1992); Tech. Adv. Mem. 93-38-002 (Sept. 24, 1993); *Frank Lyon Co. v. United State*s, 435 U.S. 561, 572-73 (1978); *Helvering v. F. & R. Lazarus & C*o., 308 U.S. 242 (1939); *Gregory v. Helvering*, 293 U.S. 465 (1935); *Comdisco, Inc. v. United States*, 756 F.2d 569, 578 (7th Cir. 1985); *Sullivan v. United States*, 618 F.2d 1001, 1006-08 (3d Cir. 1980); *Sun Oil Co. v. Commissioner*, 562 F.2d 248 (3d Cir. 1977), *cert. denied*, 436 U.S. 944 (1978); *Commissioner v. Danielson*, 378 F.2d 771, 775 (3d Cir. 1967); *Sullivan v. United States*, 461 F. Supp. 1040 (W.D. Pa. 1978); *Illinois Power Co. v. Commissioner*, 87 T.C. 1417 (1986); *Grodt & McKay Realty, Inc. v. Commissioner*, 77 T.C. 1221 (1981); *Bolger v. Commissioner*, 59 T.C. 760, 767 n.4 (1973); *Bowen v. Commissioner*, 12 T.C. 446, 459 (1949); Rev. Rul. 55-540, 1955-2 C.B. 39. Other courts have required the party requesting recharacterization to show that the parties intended a different allocation for tax purposes other than the allocation provided in the contract – see: *Balthrope v.*

*Commissioner*, 356 F.2d 28 (5th Cir. 1966); *Annabelle Candy Co. v. Commissioner*, 314 F.2d 1 (9th Cir. 1962); *Ullman v. Commissioner*, 244 F.2d 305 (2d Cir. 1959). Some courts emphasize the intention of the parties – see: *Major v. Commissioner*, 76 T.C. 239, 246 (1981); *Lazisky v. Commissioner*, 72 T.C. 495, 500-02 (1979); *G.C. Servs. Corp. v. Commissioner*, 73 T.C. 406 (1979); *Lucas v. Commissioner*, 58 T.C. 1022, 1032 (1972); *Rich Hill Ins. Agency v. Commissioner*, 58 T.C. 610, 617-19 (1972). With regard to substantive consolidation, see: *Consolidated Rock Prods. Co. v. DuBois*, 31 U.S. 510 (1940); *Reider v. FDIC* (In re Reider), 31 F.3d 1102, 1106-07 (11th Cir. 1994); *In Re Giller*, 962 F.2d 796, 798 (8th Cir. 1992); *Eastgroup Properties v. Southern Motel Assoc.*, 935 F.2d 245 (11th Cir. 1991); *In re Augie/Restivo Banking Co.*, 860 F.2d 515, 518 (2d Cir. 1988); *In re Auto-Train Corp.*, 810 F.2d 270 (D.C. Cir. 1987); *FMC Fin. Corp. v. Murphree*, 632 F.2d 413 (5th Cir. 1980); *Chemical Bank N.Y. Trust Co. v. Kheel*, 369 F.2d 845, 847 (2d Cir. 1966); *Anaconda Bldg. Materials Co. v. Newland*, 336 F.2d 624 (9th Cir. 1964); *Fish v. East*, 114 F.2d 177 (10th Cir. 1940); *Central Claims Servs. v. Eagle-Richer Indus. (In re Eagle-Richer Indus.)*, 192 B.R. 903, 905-06 (Bankr. S.D. Ohio 1996); *In re Standard Brand Paint Co.*, 154 B.R. 563 (Bankr. C.D. Cal. 1993); *In re Crown Mach. & Welding, Inc.,* 100 B.R. 24 (Bankr. D. Mont. 1989); *In re DRW Property Co.,* 54 B.R. 489 (Bankr. N.D. Tex. 1985); *In re Snider Bros.*, 18 B.R. 230 (Bankr. D. Mass 1982); *In re Vecco Constr. Indus., Inc.,* 4 B.R. 407 (Bankr. E.D. Va. 1980). Some courts have held that as a corporation becomes insolvent, its directors owe a fiduciary duty to its creditors – see: *In re Andreuccetti*, 975 F.2d 413, 421 (7th Cir. 1992); *Clarkson Co. v. Shaheen*, 660 F.2d 506, 512 (2d Cir.1 981); *In Re Kingston Square Assocs.*, No. 96B44962 (TLB), 1997 Bankr. LEXIS 1514, at *75 (Bankr. S.D.N.Y. Sept. 24, 1997); *Geyer v. Ingersoll Publications Co.*, 621 A.2d 784, 787-89 (Del. Ch. 1992); *Credit Lyonnais Bank, Nederland, N.V. v. Pathe Communications Corp.*, Civ. A. No. 12150, 1991 WL 277613, (Del. Ch. Dec. 30, 1991); *Tampa Waterworks Co. v. Wood*, 121 So. 789 (Fla. 1929); *Francis v. United Jersey Bank*, 432 A.2d 814 (N.J. 1981). On classification of lease transactions as financings or loans, also see: *United States v. Colorado Invesco, Inc.*, 902 F. Supp. 1339, 1342 (D. Colo. 1995) (*quoting In re Fabricators, Inc.*, 924 F. 2d 1458, 1469 (5th Cir. 1991)); Shawmut Bank Connecticut v. First Fidelity Bank (In re Secured Equip. Trust of Eastern Air Lines, Inc.), 38 F.3d 86, 87 (2d Cir. 1994); *In re Best Products Co.*, 157 B.R. 222, 229-30, (Bankr. S.D.N.Y. 1993); *In re Wilcox*, 201 B.R. 334, 336-37 (Bankr. N.D.N.Y. 1996). On options-to-purchase and clogging the equity-of-redemption, see: *Humble Oil & Ref. Co. v. Doerr*, 303 A.2d 898 (N.J. Super. Ch. Div. 1973); *Barr v. Granahan*, 38 N.W.2d 705 (Wis. 1949); *Getty Petroleum v. Giordano*, No. 87-3165 1988 U.S. Dist. LEXIS 4567, at *1 (D.N.J., May 19, 1988); *Blackwell Ford, Inc. v. Calhoun*, 555 N.W.2d 856 (Mich. Ct. App. 1996); *McArthur v. North Palm Beach Utils.*, 202 So. 2d 181 (Fla. 1967); *Coursey v. Fairchild*, 436 P. 2d 35 (Okla. 1967); *Hopping v. Baldridge*, 246 P. 469 (Okla. 1928); *Lincoln Mortgage Investors*, 659 P.2d at 928. The issue of preemption of state laws in the determination of whether a lease is a financing or a loan, is somewhat un-settled – see: *In re Challa*, 186 B.R. 750, 755-756 (Bankr. M.D. Fla. 1995); *In re Q-Masters, Inc.*, 135 B.R. 157, 159 (Bankr. S.D. Fla. 1991); *BFP v. Resolution Trust Corp.*, 511 U.S. 531 (1994)); *Barnhill v. Johnson*, 503 U.S. 393 (1992); *Butner v. United States,* 440 U.S. 48, 55 (1979); *MNC Commercial Corp. v. Joseph T. Ryerson & Son, Inc.*, 882 F.2d 615, 619 (2d Cir. 1989); *Morton v. National Bank*, 866 F.2d 561, 563 (2d Cir. 1989); *In re Rosenshein*, 136 B.R. 368, 372 (Bankr. S.D.N.Y. 1992); *In re Taylor*, 130 B.R. 849 (Bankr. E.D. Ark. 1991); *In re Wingspread Corp.*, 116 B.R. 915 (Bankr. S.D.N.Y. 1990) (under state law, debtors' leases were financing agreements, and not true leases); *In re Century Brass Prods., Inc.*, 95 B.R. 277, 279 (Bankr. C.D. Conn. 1989); *In Re Petroleum Products., Inc.*, 72 B.R. 739 (Bankr. D. Kan. 1987)(*affirmed*) 150 B.R. 270 (B.A.P. D. Kan. 1993); *In Re Harris Pine Mills*, 79 B.R. 919 (D. Or. 1987)(*affirmed*), 862 F.2d 217 (9th Cir. 1988); H.R. REP. NO. 95-595, at 314 (1978), reprinted in 1978 U.S.C.C.A.N. 5963, 6271; S. REP. No. 95-989, at 24 (1978), reprinted in 1978 U.S.C.C.A.N. 5787, 5812; *City of Olathe v. KAR Dev. Assocs. (In re KAR Dev. Assocs., L.P.),* 180 B.R. 629, 637 (Bankr. D. Kan. 1995)(federal law preempts). *International Trade Admin. v. Rensselaer Polytechnic Inst.*, 936 F.2d 744 transactions, c) tax consequences of sale leaseback transactions, d) disclosure and financial reporting of sale leaseback transactions.  Nwogugu (2007). [3]

---

<u>3. Leasing As A Dynamical System.</u>

The leasing process is essentially a four-stage dynamical system because: 1) the various components and relationships in the lease-system vary over time, 2) there is a clear network of relationships among distinct parties, which are defined by the lease contract, the Uniform Commercial Code, the Bankruptcy Code, custom and state laws, 3) factors that affect one component of the –lease system tend to affect other components of the system and the value of the relationships among the various components.  See: Beer (2000); Dellnitz & Junge (1999); Moore (1991); Friedman & Sandler (1996); Evans (1998); Agarwal, Bohner, O'Regan & Peterson (2002); Iacus (2001); Van Gelder (1998); Tucker (1997); Izmailov & Solodov (2001); Iri (1997); Mordukhovich & Shao (1997); Treur (2005); Hojjati, Ardabli & Hosseini (2006); Kaiser & Tumma (2004); Schultz (1997); Chehab & Lamine (2005); Sebenius (1992); Xu (2005); Vasant, Nagarajan & Yaacob (2005); Bisdorff (2000); Corbett, DeCroix & Ha (2005). The components of the system include: a) lessor, b) lessee, c) broker, d) county clerk (where leases are recorded), e) banks and financial institutions – that finance leases, f) credit enhancement vendors (eg. FGIC, FSA, etc.), g) the Lease Agreement, h) any encumberances on the subject property, i) the subject property; i) laws and regulations.  The various stages of the lease-system are as follows:

a) Stage one – the decision to lease.

---

b) Stage two – finding a tenant and negotiating and signing the lease.

c) Stage Three – performance of the lease.

d) Stage Four – any default or non-performance of lease terms, up until lease expiration.

Many existing leases are 'incomplete contracts" because they: 1) are triple-net leases, 2) have overage clauses, 3) the performance obligation is not capped/limited or clearly defined. Mooradian & Yang (2002). Gross leases are much more complete than Net-leases because they contain more specific and definite terms, and less exposure or uncertainties. Due to financial difficulties experienced by US retailers between 1995-2004, it was expected and natural that many retailer-tenants would seek to reduce the fixed portions of rents, and to increase the 'overage' or variable portions of rents. Bernfeld (Fall 2002). Grenadier (1995); Brickley (1999); McCann & Ward (2004); Tse (1999); Asabere (2004); Pretorius, Walker & Chau (2003); Seiler, Chatrath & Webb (2001); Pashigian & Gould (April 1998); Hansmann & Kraakman (2000); Mejia & Benjamin (2002). Nwogugu (2005). The effect of such 'incompleteness' in lease contracts can be substantial and depends on location, retailers' brand name, tenant marketing efforts and transaction costs (costs of re-leasing the space, litigation costs, lost sales revenues, etc.). From the retailer's/lessee's perspective, the sources of incompleteness are:

1. Operating expenses – maintenance, insurance, premises liability not covered by insurance, etc.

2. Overage rents

3. Capital expenditures

4. Premises liability

5. Natural disasters

6. Landlord's efforts in marketing the shopping mall.

7. Probability of adequate remedy for breach – suitability of pre-specified forum for resolution of disputes.

8. Lessee's Employee's effort levels at that location –

9. Lessee's intensity of utilization of space.

10. Lessee's Assignment or sub-letting rights, where Lessee must obtain lessor's permission before any assignment or sub-leasing.

11. Presence or absence of hazardous materials in the site – where lease is a NNN lease – and the extent of lessee's liability for environmental cleanups.

## 4. The Choice Between A Sale-Leaseback And New Debt.

Retailers also face the choice between a sale-leaseback and borrowing new funds. The main effect of the sale-leaseback are that: **a**) it can reduce reported assets and debt, **b**) it can increase the retailer's borrowing capacity, and can change the capital structure, anc can lower the retailer's incremental cost of capital, c) it generates cash immediately with possibly lower transaction costs (than borrowing) and at possibly higher asset values (than from borrowing), and at possibly lower implied interest rates; **d**) it can provide tax benefits, depending on whether the seller/lessee generates taxable income. The literature on wealth effects of sale-leasebacks is extensive. See: Albert & Intosh (1989); Miceli & Sirmans & Turnbull (2001); Seiler, Chatrath & Webb (2001); Petorius, Walker & Chau (2003); Graff (2001); Gibson & Barkham (2001); Mooradian & Yang (2002); Fisher (2004); Arnold (1999); Stavrovski (2004); Kangoh (1995); Ghyoost (2004); Young & Graf (1995); Garmaise & Moskowitz (2003). However, the existing literature does not analyze some of the following issues:

a) The choice between a sale-leaseback and borrowing;

b) Effect on the retailer's cost of funds;

c) Transaction costs;

d) The Retailer's probability of bankruptcy,

e) Optimal conditions for sale-leasebacks;

f) Optimal conditions for borrowing as an alternative to the sale-leaseback.

However, the structure of the sale-leaseback determines the wealth effects, if any.  The economics of leases can be modeled using fuzzy sets.  Wang & Parkan (2005); Coban & Secme (2005); Garcia, Berlanga, Molina & Davila (2004); Olson & Bayer (2003); Philpott, Hamblin, Baines & Kay (2004). Homem-De-Mello (2001).  On dynamical systems, see: Nelles (2002); Beer (2000); Dellnitz & Junge (1999); Moore (1991); Friedman & Sandler (1996); Evans (1998); Agarwal, Bohner, O'Regan & Peterson (2002); Iacus (2001); Van Gelder (1998); Tucker (1997); Izmailov & Solodov (2001); Iri (1997); Mordukhovich & Shao (1997); Treur (2005); Hojjati, Ardabli & Hosseini (2006); Kaiser & Tumma (2004); Chehab & Lamine (2005); Xu (2005); Vasant, Nagarajan & Yaacob (2005).

Let:

$L_s$ = PV of monthly lease payments under sale leaseback
$I$ = PV of monthly interest payments on loan (amortizing loan)
$S$ = sale price
$R_s$ = implicit interest rate of lease. $0 < R_s < 1$.
$R_{bb}$ = borrowing cost of the buyer/lessee before sale lease back. $0 < R_{bb} < 1$.
$R_{ba}$ = borrowing cost of the buyer/lessee after sale lease back. $0 < R_{ba} < 1$.
$R_{ts}$ = seller/lessor's tax rate.  $0 < R_{ts} < 1$.
$R_{tb}$ = buyer/lessee's tax rate.  $0 < R_{tb} < 1$.
$N$ = state – lease is an operating lease
$C$ = state – lease is a capital lease
$D$ = depreciation from property - applies to capital lease
$P$ = principal amount of loan that will be borrowed instead of sale leaseback.  This loan has monthly interest payments and same term as the sale-leaseback lease.
$R_a$ = reduction in company's borrowing cost due to lower leverage – applies only to operating lease
$R_i$ = increase in company's borrowing cost due to higher leverage from borrowing and not doing the sale leaseback
$R_r$ = firm's borrowing cost if firm borrows and does not do sale-leaseback  $0 < R_r < 1$.
$R_{sl}$ = Transaction costs if sale-leaseback, amortized over loan term, and as percentage of sale price. $0 < R_{sl} < 1$.
$R_l$ = Transaction costs if loan; amortized over loan term, and as percentage of loan principal.  $0 < R_l < 1$.
$DC$ = company's debt/capital ratio
$TC$ = total Capital
$TV$ = present value of assumed terminal value of property in sale-leasebacks classified as capital leases
$P_{dss}$ = probability of seller/lessee's bankruptcy after sale-leaseback transaction.  $0 < P_{dss} < 1$.

$P_{dsb}$ = probability of seller/lessee's bankruptcy after borrowing transaction. $0 < P_{dsb} < 1$.

$P_{dls}$ = probability of buyer/lessor's bankruptcy after sale-leaseback transaction. $0 < P_{dls} < 1$.

$P_{dlb}$ = probability of buyer/lessor's bankruptcy after borrowing transaction. $0 < P_{dlb} < 1$.

$P_t$ = probability that seller/lessee will have taxable income equal to or greater than periodic depreciation amounts. $0 < P_t < 1$.

If the retailer does a sale-lease back recorded as capital lease, its net position will be:

$N_{sl} = \{S(1-R_{sl}) + [\{(L_s * R_{ts}) + (D * R_{ts} * P_t) - L_s + R_a + TV\} * (1 - P_{dss})]\}$;

And its objective function will be:

Max $N_{sl} = \{S(1-R_{sl}) + [\{(L_s * R_{ts}) + (D * R_{ts} * P_t) - L_s + R_a + TV\} * (1 - P_{dss})]\}$.

If the retailer borrows an amount *P*, its net position will be:

$N_b = \{P(1-R_l) + [[(I * R_{ts}) - (R_i)(DC)(TC) + \{(R_{ts})(R_i)(DC)(TC)\} - I(1 - R_{ts})] * (1 - P_{dsb})]\}$;

and its objective function will be:

Max $N_b = \{P(1-R_l) + [[(I * R_{ts}) - (R_i)(DC)(TC) + \{(R_{ts})(R_i)(DC)(TC)\} - I(1 - R_{ts})] * (1 - P_{dsb})]\}$;

For the company to choose borrowing instead of the sale-leaseback, then the following conditions must exist:

1. Max$[\{R_r(1-R_{ts}) + R_l P + R_i(DC)(TC)\}, 0] < [ R_s(1-R_{ts}) - (D * R_{ts}) + R_{sl}S ]$
2. $N_b > N_{sl}$
3. Max$[\{(R_{ba} - R_{bb}) - R_{sl}\}, 0] < [R_i + R_l]$
4. $\partial R_r / \partial DC > $ Max$[(\partial R_i / \partial DC), 1]$
5. $\partial R_s / \partial S > $ Max$[(\partial R_r / \partial P), 1]$
6. $\{P(1-R_l) + [[(I * R_{ts}) - (R_i)(DC)(TC) + \{(R_{ts})(R_i)(DC)(TC)\} - I(1 - R_{ts})] * (1 - P_{dsb})]\} > $ Max$[(\{S(1-R_{sl}) + [\{(L_s * R_{ts}) + (D * R_{ts} * P_t) - L_s + R_a + TV\} * (1 - P_{dss})]\}), 0]$

The retailer/lessee will be better off doing a sale-leaseback transaction than not doing anything, if:

1. $[S - L_s - R_{sl}S - (L_s R_{ts}) - (R_a)(DC)(TC) + TV(1 - P_{dss}) ] > 0$
2. $\partial R_r / \partial DC > $ Max$[(\partial R_i / \partial DC), 1]$
3. $\partial R_a / \partial DC > $ Max$[(\partial R_i / \partial DC), 1]$
4. $R_s R_{ts} < R_{bb} R_{ts}$
5. $\partial R_{bb} / \partial DC > $ Max$[(\partial R_{ab} / \partial DC), 1]$; and $\partial^3 R_{bb} / \partial DC^3 > $ Max$[(\partial^3 R_{ab} / \partial DC^3), 1]$;
6. Max$[(\partial P_{dss} / \partial DC), 1] < (\partial R_{bb} / \partial DC)$; and Max$[(\partial^3 P_{dss} / \partial DC^3), 1] < (\partial^3 R_{bb} / \partial DC^3)$;
7. Max$[(\partial P_{dss} / \partial R_{bb}), 1] < \partial P_{dss} / \partial R_r$; and Max$[(\partial^3 P_{dss} / \partial R_{bb}^3), 1] < (\partial^3 P_{dss} / \partial R_r^3)$

The foregoing analysis applies to capital leases, but in the case of operating leases, the main differences will be:

1. There wont be any reversion of the property's terminal value to lessee.

2. There won't be any depreciation tax benefits

3. The viability of the transaction for both lessee/lessor will depend on the magnitude of the difference between the seller/lessee's and the buyer/lessor's tax rates.

5. Conclusion.

`        Leasing remains a major source of capital in the real estate sector.  Real estate constitutes a substantial portion of fixed assets (land, buildings/fixtures and lease interests), capital expenditures, loan assets and operating costs (maintenance, insurance, taxes, rents and depreciation) in many industries such as retailing, healthcare, transportation, technology, banking, oil & gas, food processing, agriculture, insurance, and lodging.  Although leases and the sale-leaseback transaction are economically viable alternatives to outright purchases (financed with debt or equity), many companies do not use real estate strategically and do not incorporate real estate strategies into their overall corporate strategy and change management processes.  The analysis of sale-leasebacks should incorporate transaction costs, bankruptcy probabilities, depreciation tax shields, the borrowing alternative, and taxes.